# Comments and Reply

## Comments on "Cooperative Density Estimation in Random Wireless Ad Hoc Networks"

Yongchang Hu, *Member, IEEE*

*Abstract*—In Onur *et al.* ["Cooperative density estimation in random wireless ad hoc networks," *IEEE Commun. Lett.*, vol. 16, no. 3, 269 pp. 331–333, Mar. 2012], two novel density estimation (DE) approaches in wireless random networks were introduced by Onur *et al.*, which are carried out respectively in cooperative and individual fashions. Both of them were derived via the maximum likelihood (ML) method. However, an implicit but fatal error was made obtaining the individual DE (I-DE) approach. This letter comments on Onur *et al.* and points out the aforementioned error. By investigating the distance order statistics (DOS) distributions in the random networks, the correct I-DE approach is presented and discussed. Simulation results also show that the correct I-DE outperforms the wrong one. More importantly, a new method that can obtain any univariate or multivariate DOS distribution is demonstrated, which is expected to be helpful for the study of the wireless communications and networking.

*Index Terms*—Random networks, distance order statistics (DOSs), density estimation, maximum likelihood estimation.

## I. Introduction

THE node density $\lambda$ is crucial to the wireless communications and networking. When $\lambda$ increases, packets surge in the wireless networks, which might jam the communication links, overload the network capacity or create large interferences at receivers. On the other hand, when $\lambda$ decreases, nodes separate from each other, which might lead to low connectivity and reachability. Therefore, the node density $\lambda$ needs to be accurately estimated for a better configuration of wireless network. In a recent paper [1], Onur et al. proposed two estimation approaches for the node density $\lambda$, the cooperative density estimation (C-DE) and the individual density estimation (I-DE), which both rely on the received signal powers collected from or shared by neighbours. To be more specific, the differences between them mainly lie in the means of collecting the samples, which are explained in Fig.1.

For a better understanding, it is necessary to briefly revisit the derivations of these two approaches. According to the *Possion point process* (PPP) [2], which is commonly used to characterize a random node deployment (due to the unknown topology of wireless *ad hoc* networks, nodes are ideally assumed to be randomly deployed), the probability of finding $k$ nodes in a bounded Borel set $\Phi \subset \mathbb{R}^m$ is given by

$$\mathbb{P}_{ppp}(\text{k nodes in } \Phi) = e^{-\lambda \mu(\Phi)} \frac{(\lambda \mu(\Phi))^k}{k!}, \quad (1)$$

where $\mu(\cdot)$ is the standard Lebesgue measure. Note that, in this paper, all formulae will be presented in an $m$-dimension space for generality. Then, some manipulations on (1) result in the distribution of the distance to its $k$-th nearest neighbour [3, eq. (2)], i.e.,

$$\mathbb{P}(r_k|\lambda) = \frac{m \, e^{-\lambda c_m r_k^m} (\lambda c_m r_k^m)^k}{r_k \Gamma(k)}, \quad (2)$$

where $\Gamma(\cdot)$ indicates the gamma function and $c_m \triangleq \frac{\pi^{m/2}}{\Gamma(m/2+1)}$ (e.g., for $m = 1, 2, 3$ there are $c_1 = 2$, $c_2 = \pi$ and $c_3 = \frac{4\pi}{3}$). Considering the radio propagation channel (the authors of [1] simply assumed the deterministic path-loss and left some more complicated channel effects for future research challenge), the $k$-th strongest received power is given by

$$P_{r,k} = CP_t \left(\frac{1}{r_k}\right)^\gamma, \quad (3)$$

where $P_t$ is the transmit power, $\gamma$ is the path-loss exponent and $C$ is the non-distance-related constant. Then, after a simple variable transformation, the probability density function (PDF) for $P_{r,k}$ can be obtained as

$$\mathbb{P}(P_{r,k}|\lambda) = \frac{m(\lambda c_m)^k \left(\frac{CP_t}{P_{r,k}}\right)^{km/\gamma} e^{-\lambda c_m \left(\frac{CP_t}{P_{r,k}}\right)^{m/\gamma}}}{\gamma P_{r,k} \Gamma(k)}. \quad (4)$$

The critical point appears in calculating the likelihood function, where an implicit fatal error was made deriving the I-DE. For the C-DE, the received power samples are collected in a distributed fashion, i.e., the sample set is $\{P_{r,k}^i | i = 1, \ldots, N\}$, where $P_{r,k}^i$ indicates the $k$-th strongest received power sample collected by the $i$-th neighbour. It is important to notice that the independence of the neighbour location guarantees that of the received power sample. Hence, if only the $c$-th strongest received power is collected, it is correct to calculate the likelihood function as

$$\mathcal{L}(\lambda | P_{r,c}^1, \ldots, P_{r,c}^N) = \sum_{i=1}^{N} ln(\mathbb{P}(P_{r,c}^i|\lambda)), \quad (5)$$

which finally leads to the unbiased ML density estimate, i.e., the C-DE, as

$$\hat{\lambda}_{C|c} = \frac{Nc - 1}{c_m \sum_{i=1}^{N} \left(\frac{P_{r,c}^i}{CP_t}\right)^{-m/\gamma}}. \quad (6)$$

Manuscript received November 10, 2015; revised February 2, 2016; accepted February 18, 2016. Date of publication February 19, 2016; date of current version April 7, 2016. This work was supported in part by the China Scholarship Council (CSC) and in part by Circuits and Systems (CAS) Group, Delft University of Technology, Delft, The Netherlands. The associate editor coordinating the review of this paper and approving it for publication was P. Serrano.

The author is with the Faculty of Electrical Engineering, Mathematics, and Computer Science, Delft University of Technology, Delft 2628 CD, The Netherlands (e-mail: Y.Hu-1@tudelft.nl; hycforever2000@gmail.com).

Digital Object Identifier 10.1109/LCOMM.2016.2532881

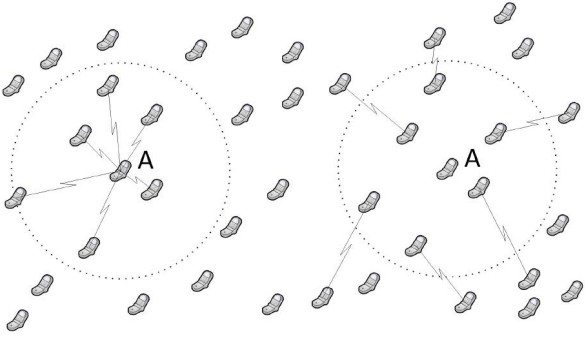

Fig. 1. The I-DE, demonstrated on the left side, locally collects the received powers from the neighbours while the C-DE, depicted on the right side, relies on the shared received powers, which are collected by the neighbours in a distributed fashion. The dotted circle indicates the transmission range of node $A$ that carries out these two approaches.

However, for the I-DE, the received power samples are locally collected, i.e., the sample set becomes $\{P_{r,i}|i=1,\ldots,N\}$. This means that the received power sample $P_{r,i}$ is correlated, since the $i$-th strongest received power also implies that there exist $i-1$ other stronger received powers. In [1], the author failed to notice this fact and wrongly calculated the likelihood function for the I-DE in the same way as for the C-DE, since the sample $P_{r,i}$ is not independent and hence $\mathbb{P}(P_{r,1},\ldots,P_{r,N}|\lambda) \neq \prod_{i=1}^{N}\mathbb{P}(P_{r,i}|\lambda)$. Therefore, constructing the likelihood function for the I-DE, i.e.,

$$\mathcal{L}(\lambda|P_{r,1},\ldots,P_{r,N}) = ln(\mathbb{P}(P_{r,1},\ldots,P_{r,N}|\lambda)), \quad (7)$$

requires the correct joint PDF $\mathbb{P}(P_{r,1},\ldots,P_{r,N}|\lambda)$, which however is very difficult at this point.

## II. MULTIVARIATE DISTANCE ORDER STATISTIC DISTRIBUTIONS IN RANDOM NETWORKS

Note that $\mathbb{P}(P_{r,1},\ldots,P_{r,N}|\lambda)$ can be easily obtained from $\mathbb{P}(r_1,r_2,\ldots,r_k|\lambda)$ after a transformation of variables, for which it is necessary to introduce the distance order statistic (DOS) distributions. In fact, the PDF in (2) is an univariate DOS distribution parametrized by the node density $\lambda$ and another one parametrized by the sample size $N$ is also presented in [4, Theorem 2.1]. However, no multivariate DOS distribution like $\mathbb{P}(r_1,r_2,\ldots,r_k|\lambda)$ has been reported before, to the best of our knowledge. Considering the univariate DOS distributions have already yielded notable influences [5]–[7], the multivariate DOS distribution might be more helpful for analysing the interference, the multi-casting, the broadcasting and etc., since the wireless communications and networking often include some one-to-many or many-to-many behaviours. In this section, besides deriving $\mathbb{P}(r_1,r_2,\ldots,r_k|\lambda)$, this section also demonstrates a new method for obtaining any multivariate DOS distribution in wireless random networks.

Assuming that every node is constrained in a given space $\Omega \subset \mathbb{R}^m$, the probability for a single node falling into an $m$-ball $\Phi \subset \Omega$ with the radius $r$ is $\mathbb{P}(\text{a single node in}\Phi|\Omega) = c_m r^m/\mu(\Omega)$. For convenience, let $\Omega$ be an $m$-ball with the radius $R$ and be concentric with $\Phi$, i.e., $\mu(\Omega) = c_m R^m$. Then, if the considered node is allowed to reside at the origin of $\Phi$ and the radius $r$ to be a nodal distance, the above probability can alternatively be viewed as the cumulative density function (CDF) for $r$, which is given by

$$\mathbb{F}(r) \triangleq \mathbb{P}(r \leq R) = r^m/R^m, \quad r \in (0, R]. \quad (8)$$

Accordingly, the PDF for $r$ can also be obtained as

$$\mathbb{P}(r) = \frac{\partial}{\partial r}\mathbb{F}(r) = mr^{m-1}/R^m, \quad r \in (0, R]. \quad (9)$$

Based on (8) and (9), the idea of obtaining the (univariate or multivariate) DOS distribution is directly applying the order statistic theory [8]. For example, the result in [4, Theorem 2.1] can easily be obtained by

$$\mathbb{P}(r_k|N) = \frac{N!}{(k-1)!(N-k)!}\mathbb{F}(r_k)^{k-1}[1-\mathbb{F}(r_k)]^{N-k}\mathbb{P}(r_k) \quad (10)$$

and its relation to (2) was also discussed in [4, Corollary 2.4]. In a similar way, it is also very convenient to obtain any multivariate DOS distribution. In this letter, for correcting the wrong I-DE, it is only required to calculate

$$\mathbb{P}(r_1,r_2,\ldots,r_k|N) = \frac{N!}{(N-k)!}[1-\mathbb{F}(r_k)]^{N-k}\prod_{i=1}^{k}\mathbb{P}(r_i)$$

$$= \frac{N!}{(N-k)!}\left(1-\frac{r_k^m}{R^m}\right)^{N-k}\prod_{i=1}^{k}\left(\frac{mr_i^{m-1}}{R^m}\right). \quad (11)$$

Finally, after introducing the node density $\lambda$ by substituting $R = \sqrt[m]{N/(\lambda c_m)}$ into (11) and taking a limit, the required joint DOS distribution can be obtained as

$$\mathbb{P}(r_1,r_2,\ldots,r_k|\lambda)$$

$$= \lim_{N\to\infty} \frac{N!}{(N-k)!}\left(1-\frac{r_k^m}{R^m}\right)^{N-k}\prod_{i=1}^{k}\left(\frac{mr_i^{m-1}}{R^m}\right)\bigg|_{R=\sqrt[m]{N/(\lambda c_m)}}$$

$$= e^{-\lambda c_m r_k^m}(m\lambda c_m)^k\prod_{i=1}^{k}(r_i^{m-1}). \quad (12)$$

## III. CORRECT INDIVIDUAL DENSITY ESTIMATION APPROACH

From (3), seeing $P_{r,i}, \forall i$ respectively as derived random variables associated with $r_i, \forall i$ and then changing the variables of (12) result in

$$\mathbb{P}(P_{r,1},P_{r,2},\ldots,P_{r,N}|\lambda)$$
$$= \mathbb{P}(r_1,r_2,\ldots,r_N|\lambda)\big|_{r_i=(\frac{P_{r,i}}{CP_t})^{-\gamma},\forall i}$$
$$\times |\det(J(P_{r,1},P_{r,2},\ldots,P_{r,N}))|$$
$$= (m\lambda c_m)^N\frac{(CP_t)^{Nm/\gamma}}{\gamma^N}e^{-\lambda c_m\left(\frac{CP_t}{P_{r,N}}\right)^{m/\gamma}}\prod_{i=1}^{N}\left(\frac{1}{P_{r,i}^{m/\gamma+1}}\right), \quad (13)$$

where $\det(\cdot)$ indicates the determinant of a matrix and the Jacobian of $[r_1,\ldots,r_N]^T$ is $J(P_{r,1},P_{r,2},\ldots,P_{r,N}) = \text{diag}([\frac{\partial r_1}{\partial P_{r,1}},\ldots,\frac{\partial r_N}{\partial P_{r,N}}]^T)$ with $\text{diag}(\cdot)$ the diagonal matrix.

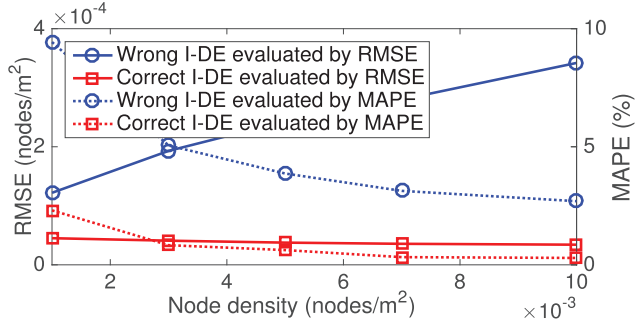

(a) First simulation when the sampling range is fixed to 100 $m$.

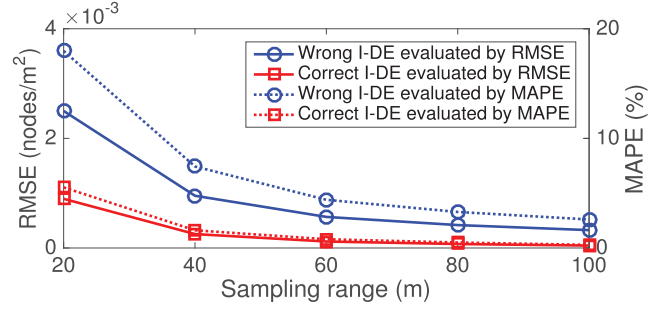

(b) Second simulation when the node density is fixed to $0.01\ nodes/m^2$.

Fig. 2. Performance of the wrong and correct I-DE: the considered node is surrounded by randomly deployed neighbors; the transmit power is set to 1 *watt* and the PLE is set to 4.

Then, plugging (13) into (7) and maximizing the likelihood function $\mathcal{L}(\lambda | P_{r,1}, \ldots, P_{r,N})$ gives a rise to an ML estimate of the density as

$$\hat{\lambda}_{ML} = \frac{N}{c_m \left(\frac{P_{r,N}}{CP_t}\right)^{-m/\gamma}}, \quad (14)$$

where an interesting observation here is that the ML solution for the I-DE is only subject to the sample from the farthest neighbor. Also note that $\hat{\lambda}_{ML}$ is asymptotically unbiased, which can be seen from its expectation as

$$\mathbb{E}[\hat{\lambda}_{ML}] = \int_0^\infty \hat{\lambda}_{ML} \mathbb{P}(P_{r,N}|\lambda)\, d(P_{r,N}). \quad (15)$$

However, directly calculating the integration in (15) would be very difficult. A simple solution is introducing an auxiliary variable $t \triangleq \lambda c_m \left(\frac{CP_t}{P_{r,N}}\right)^{m/\gamma}$ and accordingly plugging $d(P_{r,N}) = -\frac{CP_t \gamma}{mt}\left(\frac{t}{c_m \lambda}\right)^{-\frac{\gamma}{m}} d(t)$ into (15), which results in

$$\mathbb{E}[\hat{\lambda}_{ML}] = \frac{\lambda N}{\Gamma(N)} \int_0^\infty t^{(N-1)-1} e^{-t}\, d(t) = \frac{N}{N-1}\lambda, \quad (16)$$

where $\Gamma(x) = (x-1)! = \int_0^\infty t^{x-1} e^{-t}\, d(t)$. Here, it is obvious that $\mathbb{E}[\hat{\lambda}_{ML}] = \lambda$ when $N \to \infty$, which indicates that $\hat{\lambda}_{ML}$ is asymptotically unbiased.

Finally, for an absolutely unbiased ML solution, the correct I-DE is obtained as

$$\hat{\lambda}_I = \frac{N-1}{N}\hat{\lambda}_{ML} = \frac{N-1}{c_m \left(\frac{P_{r,N}}{CP_t}\right)^{-m/\gamma}}. \quad (17)$$

## IV. NUMERICAL RESULTS

Two Monte Carlo simulations have been conducted to evaluate the correct I-DE, where the wrong I-DE of [1] is also considered for comparison. Note that the node density is calculated as $\lambda \triangleq \frac{\text{sample size}}{\text{sampling area}}$ by definition, where the sampling area for the I-DE is within the transmission range, i.e., the sampling range. The impacts of the node density and the sample range are investigated respectively, since, in practice, the node density is one of the most significant parameters of wireless network and the sampling range is determined by the configuration for wireless communications.

Two measures of estimation accuracy are computed for evaluation: mean absolute percentage error (MAPE) and root mean square error (RMSE). Note that former one was also called average absolute percentage deviation (AAPD) in [1]. The use of these two measures is aimed at providing more thorough and insightful understandings of density estimation.

*a) First Simulation:* This simulation studies the impact of the node density with a fixed sampling range. Note that, in this case, a large value of the node density also implies a large sample size. As shown in Fig. 2a, the correct I-DE remarkably outperforms the wrong I-DE. More interestingly, with more samples, the correct I-DE yields smaller values of both MAPE and RMSE while the wrong I-DE yields a smaller value of MAPE but a larger value of RMSE.

To explain that, first notice the ML property implies, with more samples, the nodal distance to the farthest neighbor becomes closer to the sampling range in random networks. To be explicit, denoting the sampling range as $R_{range}$, the nodal distance $r_N$ to the farthest neighbor approaches $R_{range}$ ($r_N < R_{range}$) when $N \to \infty$, and then it is easy to observe the ML sense of our correct I-DE since

$$\lim_{N \to \infty} \hat{\lambda}_I \bigg|_{P_{r,N}=CP_t\left(\frac{1}{r_N}\right)^\gamma} = \lambda \lim_{N \to \infty} \frac{R_{range}^m}{r_N^m} = \lambda, \quad (18)$$

which also reveals another interesting fact that $\hat{\lambda}_I$ always overshoots the true value. Therefore, when the sample size increases, a decreasing estimation error results in smaller values of RMSE and MAPE for the correct I-DE. However, for the wrong I-DE, it is neither an ML nor an unbiased solution, which is readily seen from the expectation obtained using the joint PDF in (13). To save space, we will not further discuss it in details, but only observe the estimation bias from the simulation results. With an increasing sample size, the estimation bias of the wrong I-DE grows rather severer, thus yielding a large value of RMSE. The MAPE, on the other hand, actually presents the average ratio of the estimation error (including the bias and the deviation to the mean) to the true value of the node density. Considering that node density might increase relatively faster, it is now understandable that the value of MAPE can still remain a deceasing tendency with an increasing estimation error.

*b) Second Simulation:* This simulation investigates the impact of the sampling range with a fixed node density, such

that the MAPE is not impacted by the node density change. As shown in Fig. 2b, the correct I-DE is again better than the wrong one. Moreover, more geometric knowledge is obtained with a large sampling range, hence reducing the estimation error. Also, note that, limited by the transmission range, the I-DE merely estimates the so-called local density, however the C-DE has a larger sampling area owing to the sample sharing, thus yielding a better performance than the I-DE. This is also the same reason why the C-DE performs better with a large $c$.

## V. Conclusions and Discussions

In [1], the authors overlooked the fact that the order statistics from the same rank are dependent to each other and hence the joint PDF of order statistics is not a simple product of the PDFs of single order statistic. This directly led to the wrong ML method for the I-DE.

In order to correct the wrong I-DE, this letter has firstly demonstrated a new method that can obtain any multivariate DOS distribution, which might be useful for analysing the wireless communications and networking, since most behaviours in the wireless networks are in forms of one-to-many or many-to-many such as the multi-casting and the broadcasting.

Then, the correct unbiased I-DE has been presented in (17). The numerical results have shown that the correct I-DE outperforms the wrong one. The impacts of the node density and the sampling range are also investigated and discussed.

Furthermore, for a better understanding of the density estimation, some significant observations about the correct I-DE and the C-DE are summarized as follows:

1) The I-DE locally collects the samples by the considered node while the C-DE requires all the neighbors to collect the samples in a distributed fashion and to share them with the considered node.
2) To facilitate the C-DE, the sample should be piggybacked on the beacon message or some other packets that flood in the networks, which might require the assistance from the network layer or the MAC layer. The I-DE, on the other hand, is more convenient, which can individually estimate the density without any external help.
3) Due to the sampling range, the I-DE estimates the local density around the considered node while the C-DE estimates the density in a larger scale.
4) The derivation of the correct I-DE considers all the locally collected samples. Interestingly though, the ML solution for the I-DE is only subject to the sample from the farthest neighbours.
5) The C-DE, which is also an ML solution, does not require all the samples to be collected from the farthest neighbours, since the geometric independence of node already guarantees that of the collected sample.

Finally, to end this letter, some issues for practical implementation of the density estimation should be addressed. Note that the received power sample is assumed to be only subject to the geometric path-loss in this letter. However, the sample can be obtained after successful segregation of complicated channel effects. To be specific, this segregation can be achieved by smart receiver designs or signal processing techniques. For example, [9] introduces some methods to estimate the desired received power sample, which acts as one of the parameters in those channel effects. Admittedly, in a severe environment, the segregation might be unreliable and hence a new density estimation method, which considers more complicated channel effects, is required. However, it is already beyond the scope of this letter and left for the future work.

## Acknowledgment

The author would like to sincerely thank the generous help from prof.dr Ertan Onur, a member of IEEE and ACM (e-mail: onure@ieee.org) and prof.dr.ir. Geert Leus, a Fellow of IEEE (e-mail: G.J.T.Leus@tudelft.nl). The author appreciates very much for the anonymous reviewers and their constructive suggestions.